\documentclass[aps,nofootinbib,twocolumn]{revtex4}
\usepackage{graphicx,amsmath,amssymb,amsbsy,latexsym,verbatim}
\usepackage{enumerate}

\newcommand{\be}{\nopagebreak[3]\begin{equation}}
\newcommand{\ee}{\end{equation}}
\newcommand{\ba}{\nopagebreak[3]\begin{eqnarray}}
\newcommand{\ea}{\end{eqnarray}}

\newcommand{\bc}{}

\begin{document}
\title{ \Large Why all these prejudices against a constant?}
     \author{Eugenio Bianchi, Carlo Rovelli}
     \affiliation{Centre de Physique Th\'eorique de Luminy\footnote{Unit\'e mixte de recherche (UMR 6207) du CNRS et des Universit\'es de Provence (Aix-Marseille I), de la M\'editerran\'ee (Aix-Marseille II) et du Sud (Toulon-Var); laboratoire affili\'e \`a la FRUMAM (FR 2291).}, Case 907, F-13288 Marseille, EU}
\date{\small \today}
\begin{abstract}
\noindent
The expansion of the observed universe appears to be accelerating. A simple explanation of this phenomenon is provided by the non-vanishing of the cosmological constant in the Einstein equations.  Arguments are commonly presented to the effect that this simple explanation is not viable or not sufficient, and therefore we are facing the ``great mystery" of the ``nature of a dark energy". We argue that these arguments are unconvincing, or ill-founded.  \end{abstract}

\maketitle
\section{Introduction}

{\em ``Arguably the greatest mystery of humanity today is the prospect that 75\% of the universe is made up of a substance known as `dark energy' about which we have almost no knowledge at all."}  

This is the opening sentence of a (good) popularization article on dark energy \cite{PW}.  It is just an example, out of very many that can be found in the popular-science and in the technical \cite{rev,joe} literature, of how the `dark-energy' issue is perceived by many scientists, and presented to the large public.   

We argue here that there is something scientifically very wrong in this presentation. There is no ``great mystery", in an accelerated expansion of the universe.  This is a phenomenon which is clearly predicted and simply described by well-understood current physical theory. It is well understood in the context of general relativity, which naturally includes a cosmological constant.  We argue below that the common theoretical objections against this interpretation of the acceleration are either weak, or ill-founded.

The Lambda-Cold Dark Matter  ($\Lambda$CDM) model, which is today the standard model in cosmology, assumes the presence of the cosmological term in the Einstein's equations. This model is ``almost universally
accepted by cosmologists as the best description of the present data" \cite{data}. What we say here does not mean that there is no interest in exploring theoretical explanations of the acceleration \emph{alternative} to the $\Lambda$CDM model. Good science demands us to be \emph{a priori} skeptical of any theory, even when it works well, and always explore alternatives.  Even less are our observations criticisms to the observational work aiming at {\em testing} the $\Lambda$CDM scenario.  Exploring alternative theoretical explanations, and pushing the empirical test of \emph{all} the theories we have, is obviously good science.    

But what we say \emph{does} mean that it is misleading to talk about ``a mystery" (not to mention ``the greatest mystery of humanity"), for a phenomenon that has a current simple and well-understood explanation within current physical theories.  

It is especially wrong to talk about a mysterious ``substance" to denote dark energy. The expression ``substance" is inappropriate and misleading. It is like saying that the centrifugal force that pushes out from a merry-go-round is the  ``effect of a mysterious substance". 

There are three stories (of very different kind) that are routinely told in presenting the difficulties of the cosmological constant scenario.  These are: \emph{i.} The alleged historical rejection of the cosmological constant by Einstein, and then by the general-relativity community.  \emph{ii.} The coincidence problem. \emph{iii.} The enormous difference between the small value of the cosmological constant revealed by the cosmic acceleration and the large value that can be derived from quantum field theory.  We believe that there is confusion, either historical or conceptual, in {\em each} one of these three stories, as commonly presented, and we discuss them below.  

There is probably nothing very original in this note. The points we make here can be heard in discussions among physicists.  However, for some reason they do not have much space in the dark-energy literature. We though it appropriate to make them available in writing. 

\section{Einstein greatest blunder}

As well known,  in the context of his seminal cosmological work \cite{ec}, Einstein considered the possibility of adding the cosmological term to his field equations
\ba
R_{\mu\nu}-\frac12 R g_{\mu\nu}= 8\pi G\, T_{\mu\nu}\hspace{1em}& \to& \nonumber\\
&&{\ } \hspace{-10em}
R_{\mu\nu}-\frac12 Rg_{\mu\nu} +\lambda g_{\mu\nu}= 8\pi G\,T_{\mu\nu},
\label{eq}
\ea
but then referred to this step as his ``greatest blunder" \cite{gamov}.  Why should have Einstein considered the addition of the $\lambda$ term such a great blunder?  

The story often told is that Einstein was unhappy with this term since it allegedly spoiled the beauty of the theory. The cosmological term in the Einstein equations is often called ``infamous", and said to ruin the simplicity of general relativity.  
This is nonsense, and has nothing to do with Einstein's reason to talk of a ``greatest blunder".  
Between 1912 and 1915, Einstein had published all sort of \emph{wrong} equations for the gravitational field (see for instance \cite{ew}) -- not to mention articles that are mathematically wrong, as Einstein himself later conceded (for instance the claim that DeSitter universe must contain matter  \cite{eds}). He later corrected his wrong equations and wrong conclusions, and never thought that a physically wrong published equation should be thought as a ``great blunder".  Why then a great blunder, even the ``greatest blunder" with the cosmological constant? 

The reason is that --with hindsight-- Einstein had been really stupid in his --otherwise spectacular-- early cosmological work.  His general-relativity equations imply that the universe cannot be static. Had he been more courageous, he could have easily \emph{predicted} that the universe is not static: it either expands or contracts.  He could have \emph{predicted} in 1917 the expansion of the universe, for which the first hints of empirical evidence were just emerging, and which was then commonly accepted only after Hubble's 1929 work.  But Einstein did not find the courage to believe his own theory and made a very unnatural effort to force the theory into being consistent with the the static character of the universe that was commonly assumed at the time. The man who had the courage to tell everybody that their ideas on space and time had to be changed, then did not dare predicting an expanding universe, even if his jewel theory was saying so. 

But Einstein blunder was worse than this, and truly ``great": in order to force his theory into being consistent with stability, the key point was not so much adding the $\lambda$ term, but rather the rude hypothesis that the value of $\lambda$ {\em exactly} matched the energy density $\rho$ of matter (and this exactly matched the radius of the universe), so to have equilibrium between the attractive (conventional) gravitational force and the repulsive cosmological-constant force. For a closed homogeneous and isotropic universe of radius $a$, the equations (\ref{eq}) reduce to the Friedmann equation
\be
\left(\frac{\dot a}{a}\right)^2\equiv H^2=\frac{8\pi G}{3}\rho+\frac{\lambda}{3}-\frac{1}{a^2}
\label{fr}
\ee
and to have a static universe, namely $\dot a \equiv da/dt=0$, Einstein demanded
\be
\lambda={4\pi G}\rho=a^{-2}.
   \label{equi}
   \ee
This exact balance is absurdly \emph{ad hoc}. But this is not yet the worse. The worse part of the story is that such an effort was bad mathematics and bad physics anyway. For ordinary matter $\rho$ scales with $a$ as $\rho\sim a^{-3}$. Taking the time derivative of the previous equation, and using (\ref{equi}), one finds easily for $a=a_0+\delta a$
\be
\frac1a\frac{d^2a}{dt^2}=\frac{1}{3}(\lambda-4\pi G\rho)\sim (4\pi G \rho) \delta a;
\ee
that is, the resulting equilibrium is \emph{unstable}.  This means that Einstein effort to force the theory into predicting a stable static universe was misguided anyway, with or without cosmological constant!  Einstein missed this instability, which any graduate student could check today in a few lines (it was pointed out by Eddington a few years later \cite{eddington}).

Einstein had in his hands a theory that predicted the cosmic expansion (or contraction) {\em without} cosmological constant, {\em with} a generic value of the cosmological constant, and even, because of the instability, {\em with a fine-tuned value} of the cosmological constant.  But he nevertheless chose to believe in the fine-tuned value, goofed-out on the instability, and wrote a paper claiming that his equations were compatible with a static universe!  These are facts.  No surprise that later he referred to all this as his ``greatest blunder":  he had a spectacular prediction in his notebook and contrived his own theory to the point of making a mistake about stability, just to avoid making ... a \emph{correct} prediction!  Even a total genius can be silly, at times.

Why is this relevant for the debate about the cosmological constant?  Because short-cutting this story into reporting that Einstein added the cosmological term and then declared this his ``greatest blunder" is to charge the cosmological term with a negative judgment that Einstein certainly never meant.

In fact, it may not even be true that Einstein {\em introduced} the $\lambda$ term because of cosmology. He probably knew about this term in the gravitational equations much earlier than his cosmological work. This can be deduced from a footnote of his 1916 main work on general relativity \cite{foot} (the footnote is on page 180 of the English version). Einstein derives the gravitational field equations from a list of physical requirements. In the footnote, he notices that the field equations he writes are {\em not} the most general possible ones, because there are other possible terms.  The cosmological term is one of these (the notation ``$\lambda$" already appears in this footnote). 

The most general low-energy second order action for the gravitational field, invariant under the relevant symmetry (diffeomorphisms) is
\be
S[g]=\frac{1}{16\pi G}\int (R[g]-2\lambda)\sqrt{g}, 
\ee
which leads to (\ref{eq}). It depends on \emph{two} constants, the Newton constant $G$ and the cosmological constant $\lambda$, and there is no physical reason for discarding the second term.  

From the point of view of classical general relativity, the presence of the cosmological term is natural and a vanishing value for  $\lambda$ would be more puzzling than a finite value: the theory naturally depends on two constants; the fact that some old textbooks only stress one ($G$) is only due to the fact that the effects of the second ($\lambda$) had not been observed yet.
 
In gravitational physics there is nothing mysterious in the cosmological constant. At least nothing \emph{more} mysterious than the Maxwell equations, the Yang-Mills equations, the Dirac equation, or the Standard Model equations. These equations contain constants whose values we are not able to compute from first principles. The cosmological constant is in no sense more of a ``mystery" than any other among the numerous constants in our fundamental theories. 

\section{Cosmic coincidence}

Suppose that the history of the universe is well described by the Einstein equations with cosmological constant.  Then there is something curious going on.  On a generic cosmological solutions, the Friedmann equation reads
\begin{equation}
H^2=\frac{8\pi G}{3}\rho+\frac{\lambda}{3}-\frac{k}{a^2}
\end{equation}
where $k=1,0,-1$ for closed, spatially flat, and non-flat open universes, respectively (the $k=1$ case is (\ref{fr}).) To compare the size of the three terms in the right hand side of this equation it is custom to divide the equation by the left hand side and define 
\be
\Omega_{m}=\frac{8\pi G\rho}{3H^2},\ \ \ \ \  \Omega_\Lambda=\frac{\lambda}{3H^2}
,\ \ \ \ \  \Omega_{k}=-\frac{k}{a^2 H^2},
\label{omega}
\ee
so that
\be
\Omega_{m}+\Omega_\Lambda+\Omega_{k}=1.
\ee

Let us open a parenthesis about $\Omega_{k}$ first. Observation appears to point towards $\Omega_{m}+\Omega_\Lambda\sim1$ within a few percent, and therefore constrain $\Omega_{k}$  to be small compared to unity. See Figure 1. This means that
the universe is approximately flat on the scale of our horizon.  More precisely, (from the last equation in (\ref{omega})) the radius of the spatial universe ($a$) is at least about one order of magnitude larger that the radius of the visible universe ($\sim H^{-1}$).

In a very analogous situation, humanity made a well known mistake: from the observation that the Earth is approximately flat within the visible horizon, it was concluded that the Earth is flat. 

So, we believe that perhaps one should not jump to the conclusion that the spacial universe is flat either, as it is often done.  (On this, see also \cite{ellis}.) Our own prejudice is in favor of a large but closed universe (for the same reasons for which a flat infinite Earth is conceptually unpalatable), and one may want to read this in the data (look at Figure 1 with more attention\footnote{Reference \cite{ellis} gives $\Omega_{m}+\Omega_\Lambda=1.054^{+0.048}_{-0.041}$ implying $k = +1$.}). Whatever the prejudice, observational evidence is consistent with $k=0$ but is far from selecting out {\em only} this possibility.
\begin{figure}[h]
\centering
\includegraphics[scale=.7]{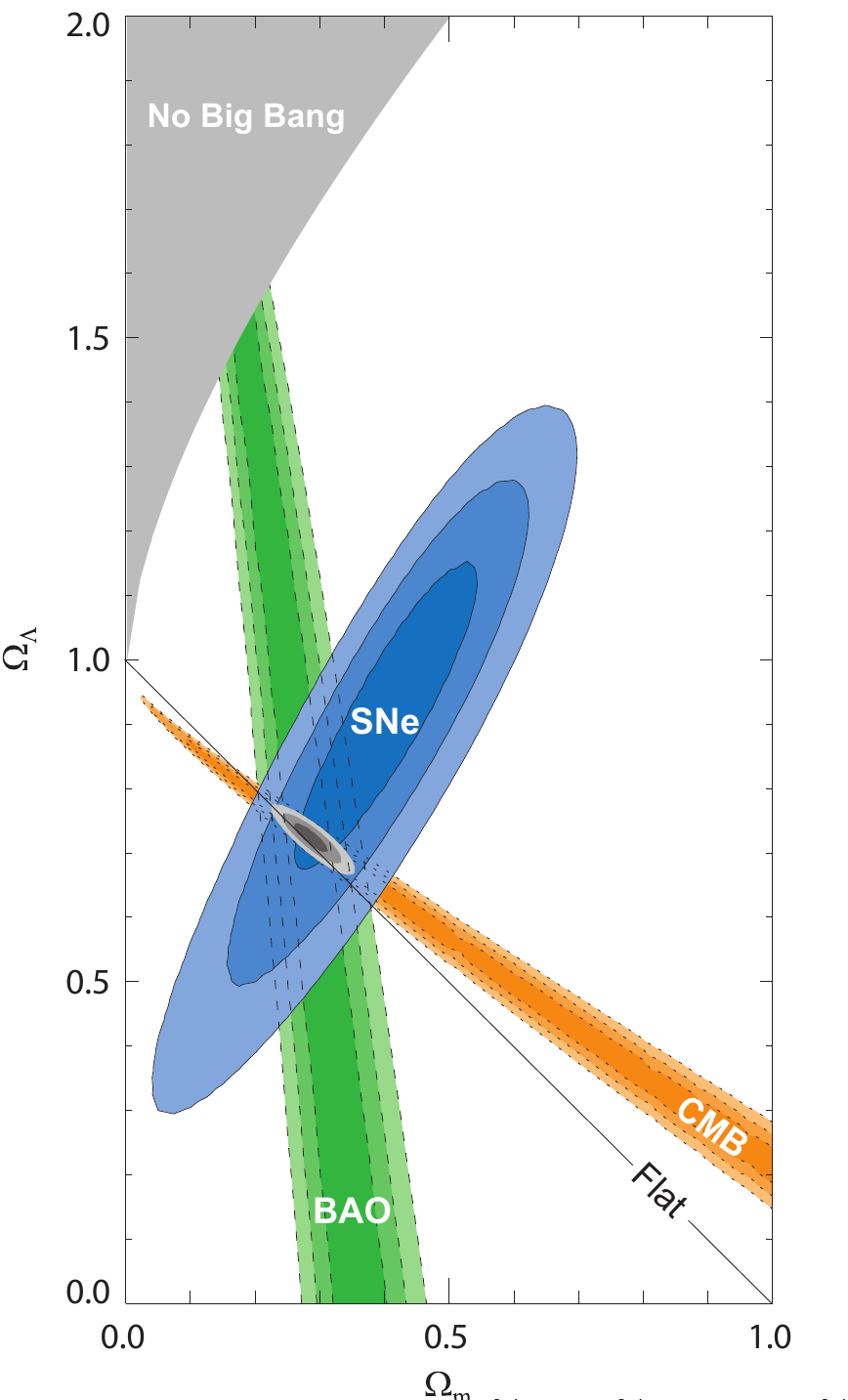}
\caption{Confidence level contours of 68.3\%, 95.4\% and 99.7\% in the
$\Omega_\Lambda-\Omega_m$ plane from the Cosmic Microwave Background, Baryonic Acoustic Oscillations and the Union SNe Ia set, as well as their combination \cite{k}. A value above the ``Flat" line indicates a closed universe.}
\end{figure}

For the sake of the present discussion, however, let us assume that $\Omega_{k}=0$, close the parenthesis, and consider the other two parameters.  According to the $\lambda$CDM model, 
\begin{equation}
\Omega_{m}=\Omega_{b}+\Omega_{dark\ matter}
\end{equation}
and observation gives  \cite{data}
\begin{eqnarray}
\Omega_{b}&\sim& 0.0227\pm0.0006, \nonumber\\
\Omega_\lambda&\sim& 0. 74\pm 0.03.   \label{tn}
\end{eqnarray}
Hence ``dark energy" is roughly 30 times bigger that the observed matter energy (and 2.5 times the energy density that includes dark matter). These are observed facts.  

Now the standard ``coincidence" argument against the credibility of this scenario is the following. 
Since matter density scales with the third power of the inverse scale factor for ordinary baryonic matter (dust), 
\begin{equation}
\rho_{b}(a) \sim 0.0227\ a^{-3} H^2_o
\end{equation}
(where $H_o$ is the present value of $H$) while $\rho_\Lambda(a)=\lambda/3$ is independent from $a$, the dynamics of a generic expanding universe will be first dominated by the matter term $\rho_{b}$, and later be dominated by the cosmological constant term $\rho_\Lambda$. Only in a ``brief" intermediate phase will the two terms be of the same order of magnitude.  But (\ref{tn}) indicates that we are precisely in this intermediate phase.  This is a strange ``coincidence" that indicates that we live in a peculiar moment of the history of the universe. Say we want to hold a ``cosmological principle" stating that we are not in a special place in the universe, in space or in time. Thus, there is a contradiction between the $\Lambda$CDM model and such a cosmological principle: to believe that the observed acceleration is caused by a cosmological term in Einstein equations requires us to believe also that we are in a very special moment of the history of the universe.  This is the ``coincidence argument" against the cosmological constant scenario. 

Probability arguments are often tricky and should be handled with care.  We think the argument above is incorrect, for two reasons.  

First, if the universe expands forever, as in the standard $\Lambda$CDM model, then we cannot assume that we are in a random moment of the history of the universe, because all moments are ``at the beginning" of a  forever-lasting time. 

Hence we can only reasonably ask whether or not we are in a special point in a time span of the same order of magnitude of our own cosmic time. Let us, say, triple the current age $t_H$ of the universe, and ask whether we are in a special moment in the history of the universe between $t=0$ and $t=3 t_H$. Common plots in cosmology appear to indicate so, but these plots are usually in logarithmic scale. Why should we use equiprobability on a logarithmic scale?  More reasonable is to use equiprobability in proper time, or in the size of the scale factor $a$.  Let's choose the second (the conclusion is not affected). Let's plot the values of $\rho_b$ and $\rho_\Lambda$ as they evolve with $a$. These are given in Figure 2 as functions of $a$ and in Figure 3 as functions of $\ln{a}$. At present, the ratio of the two is more than one order of magnitude and less than two orders of magnitudes.  Consider the region where the two densities are within two orders of magnitude from each other. This is the interval of the $a$ axis in which the dotted curve is within the grey band, namely the  interval of the $a$ axis between the two dotted vertical lines. This interval includes more than half of the total interval in $a$ considered. 
\begin{figure}[h]
\centering
\includegraphics[scale=.7]{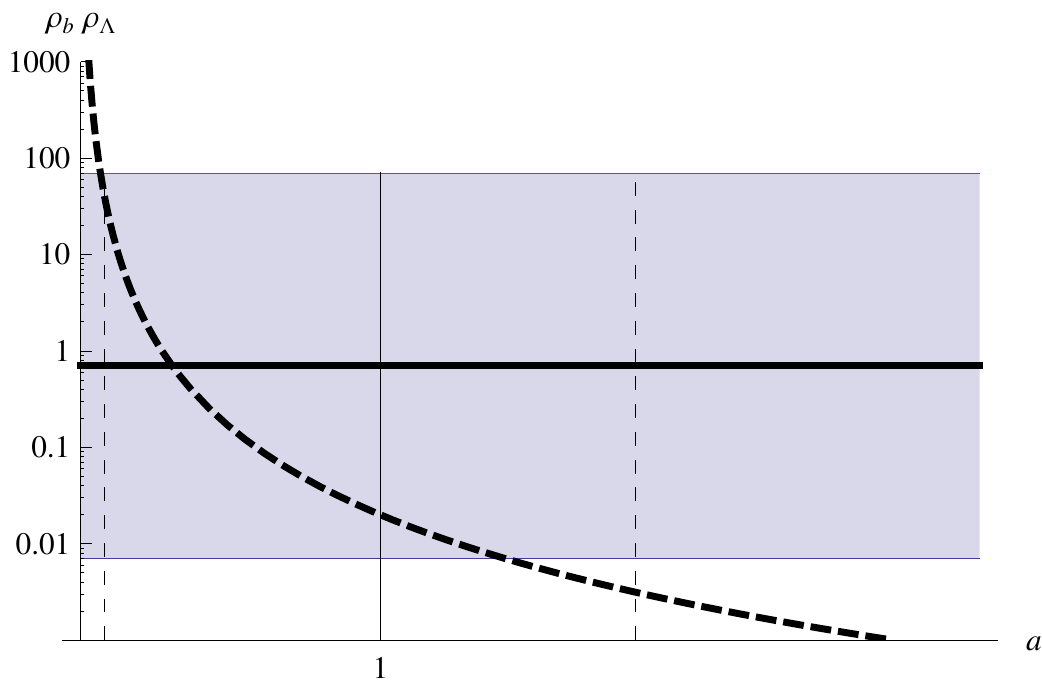}
\caption{Dark energy (continuous line) and baryon energy (dashed line) as a function of the size $a$ of the Universe. The two are within two orders of magnitudes from one another (as they are today ($a=a(t_H)=1$), for all the interval included between the two vertical dashed lines, namely for the major part of the history of the universe, on our time scale.}
\end{figure}
\begin{figure}[h]
\centering
\includegraphics[scale=.7]{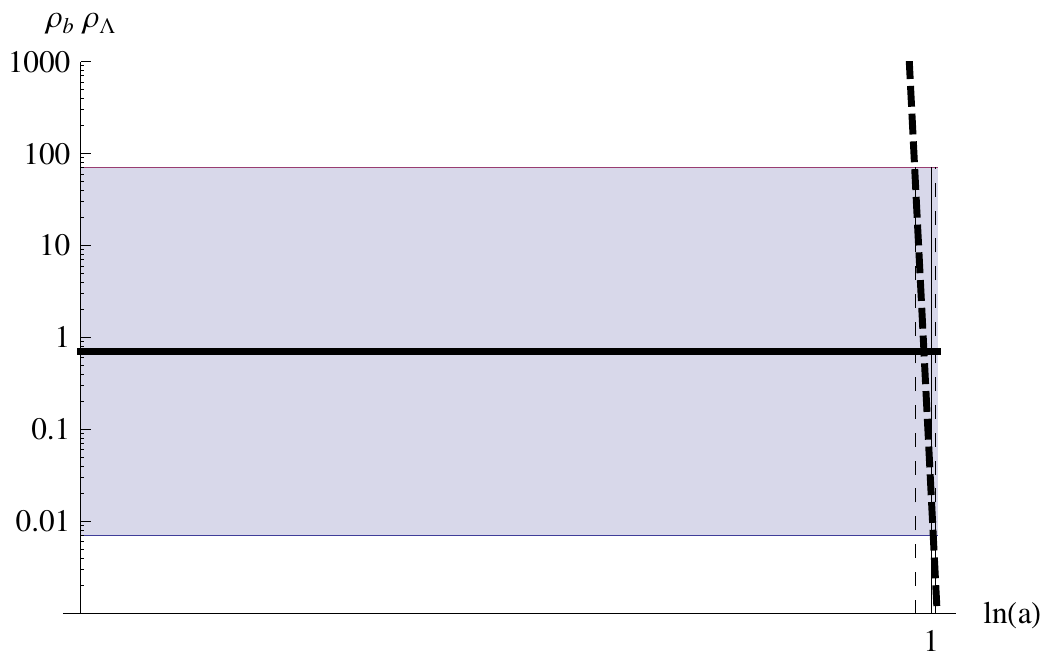}
\caption{Same as Figure 2, but in a logarithmic scale for the scale factor (from Planck to cosmic scale).}
\end{figure}

From this point of view, the ``coincidence problem" is put back into perspective. It is not a problem of a fine-tuned coincidence. It is rather an issue of order of magnitudes: why is the order of magnitude $t_H$ comparable with the one determined by $\lambda$? That is, applying the cosmological principle to orders of magnitude, why do we happen to live in an age of the universe which is not many orders of magnitudes smaller or larger?

Here comes the second objection to the coincidence argument: the cosmological principle cannot be applied uncritically, as was pointed out by Robert Dicke \cite{dicke} in 1961 and convincingly argued by Brandon Carter \cite{carter} in 1973.  For instance, a rigorous application of the cosmological principle would lead us to expect that the density around us must be of the same order of magnitude as the average density of the universe (which is manifestly false); or, to put it visually, that the Earth is mostly covered by land and not oceans (most humans observe land and not water around them.)  Humans do not live in a random location on Earth. They leave on land and not on water.  Our civilization is not located in a random location in the universe: it is located on a very high peak on the density fluctuations, far out of statistics.   This observation is a very mild form of anthropic principle. This is a form of anthropic principle that cannot be rejected even by those (like us) who most furiously oppose the use of anthropic-principle arguments in theoretical physics.  In order for us to comfortably exist and observe it, we must be on land and not on water, inside a galaxy and not in intergalactic space, and in a period of the history of the universe where a civilization like ours could exist. Namely when heavy elements abound, but the universe is not dead yet. Since the universe evolves, this period is not too long, on a logarithmic scale: it cuts out scales much shorter or much longer than ours.

To summarize, in a universe with the of value of $\lambda$ like in our universe, it is quite reasonable that humans exist during those 10 or so billions years when $\Omega_b$ and $\Omega_\lambda$ are within a few orders of magnitude from each other. Not much earlier, when there is nothing like the Earth, not much later when stars like ours will be dying. Of course there is nothing rigorous in these arguments. But this is precisely the point: there is nothing rigorous or convincing in the coincidence argument.

One sits through many seminars in which the speaker motivates his search for an alternative to the cosmological constant scenario with the coincidence argument... and then discusses an alternative suffering precisely for the same alleged difficulty!  

The coincidence argument is very weak: it cannot be applied towards the future because of the infinity of future in the current scenario; it fails towards the past unless one assumes equi-probability on a logarithmic scale, which has no ground; and is based on a version of the cosmological principle which is known to fail on many other situations (such as the density of the universe).

\section{The vacuum energy in quantum field theory} 

Finally, let us come to the argument that many considered the strongest against the  cosmological constant scenario: the quartic divergence of the vacuum energy density in  quantum field theory (QFT). To begin with, here is a naive version of the argument:  a formal quantization of a field theory leads to a divergent energy for the vacuum state.   If we assume that divergences are controlled by a physical cut-off, say at the Planck scale $M_P$, then the theory predicts a Planck-scale vacuum energy.  This behaves like an effective cosmological constant. Therefore one can say that QFT {\em predicts} the existence of a cosmological constant. This sounds good. However, the predicted cosmological constant 
has a value 
\begin{equation}
\lambda_{\rm\scriptscriptstyle  QFT}\sim c^4M^2_P/\hbar^2\sim10^{87}s^{-2},
\label{lscal}
\end{equation}
which is about 120 orders of magnitude larger that the observed one: 
\begin{equation}
\lambda \sim10^{-35}s^{-2}. 
\end{equation}
Thus, the cosmological constant is predicted by particle physics, but the prediction is quantitatively wrong by 120 orders of magnitude  (``The worst theoretical prediction in the history of physics" \cite{worse}).

Here is a more refined version of the argument, which does not require talking about infinities.  A given QFT with a finite cut-off $M$ can be interpreted as an {\em effective} theory, valid at energy scales well below $M$, obtained from a more complete, high-energy theory, by integrating away the high-energy modes.  If we start from the high-energy theory and integrate the high-energy modes, the physical constants in the theory get renormalized by the radiative corrections due to the high-energy modes, and scale. In general, a dimensionful constant will be taken to the scale $M$ by this process, unless it is protected by some mechanism (such as a symmetry). If such mechanism is at work, then the dimensionful constant will scale at most logarithmically (hence not much) between the scale $M$ and low energy. In principle, it could still be small at low energy, if it happens that its bare value at high energy exactly --or almost exactly-- compensates the radiative corrections; but this would demand an ``unnatural" coincidence between bare value and radiative corrections. Hence it is reasonable to postulate a ``naturalness principle", and expect that no low-energy constant scales polynomially unless protected. Thus,  in the low-energy theory naturalness implies that we do not expect to see a constant that scales quadratically (like a mass) --or, worse, quartically-- unless some specific mechanism like a symmetry protects it \cite{tHooft:1979bh}. 

The classical action that describe the world at the best of our current knowledge is the action
\begin{eqnarray}
S[\phi,g]&=&S_{\rm SM}[\phi,g]+\frac{1}{16\pi G}\int (R[g]+\lambda)\sqrt{g}
\label{action}
\end{eqnarray}
where we have collectively denoted $\phi$ the ensemble of the ``matter" (that is: fermion, gauge and Higgs) fields and $S_{\rm SM}$ is the Standard Model lagrangian written on a curved spacetime. To study the corresponding quantum theory, we can expand the fields around a vacuum solution 
\begin{eqnarray}
\phi &=& \phi_0+\delta\phi,  \label{exp}\\
g&=& \eta+h,  \label{exp2}
\end{eqnarray}
where $\eta$ is the Minkowski metric, and compute perturbatively the effective action 
\begin{eqnarray}
\Gamma [\phi,g]&=& S [\phi,g] + \hbar \Gamma_1[\phi,g]+...
\end{eqnarray}
which is the Legendre transform of $E[j_\phi,j_g]$  formally defined as
\begin{eqnarray}
e^{iE[j_\phi,j_g]}&=&\int D[\phi] D[g]\  e^{iS[\phi,g]+i\!\int\! \phi j_\phi+i\!\int\! g j_g}.
\label{pi}
\end{eqnarray}
Already at one loop, we obtain a term 
\begin{eqnarray}
\Gamma [\phi,g]&=& S [\phi,g] +\int \Lambda(M) \, \sqrt{g} + ...
\end{eqnarray}
which on purely dimensional grounds will contain contributions of order 
\begin{eqnarray}
\Lambda(M)=O(M^4)
\end{eqnarray}
in the cut-off $M$. This term renormalizes $\lambda$ in (\ref{action})
\begin{eqnarray}
\frac{\lambda}{16\pi G}\longrightarrow 
\frac{\lambda}{16\pi G} + \Lambda(M)\equiv \frac{\lambda_{\rm\scriptscriptstyle  QFT}}{16\pi G}
\label{scaling}
\end{eqnarray}
(notice that there is no Planck constant in $\frac{\lambda}{16\pi G}$, which is a classical constant; the Planck constant enters in $\Lambda(M)$.) If we take a cut-off of the order of the Planck mass $M_P$, we have 
\begin{equation}
 \Lambda(M)\sim M_P^4
\end{equation}
which gives (\ref{lscal}); but we do not need to assume that the theory is valid up to the Planck mass to find the problem. It is sufficient to cut-off the theory, for instance, at around $M=M_{100\, GeV}\sim100 GeV$, where we know electrons are still point-like, and calculate electron-loops contributions to $\Lambda$. These are still 55 orders of magnitude larger than the observed $\lambda$.

No fully convincing mechanism to protect $\lambda$ from the scaling (\ref{scaling}) is currently known; therefore QFT indicates that $\lambda$ should take a value $\lambda_{\rm\scriptscriptstyle QFT}$ of the order of the cut-off.  Naturalness --which has often proven fruitful in particle physics, yielding remarkable predictions-- indicates that in the real world there should be something that we have not yet understood, preventing $\lambda$ from scaling so much. 

The problem is similar (but worse) to the problem given by the Higgs mass, which scales quadratically in the standard model, and can be taken as an indication that ``there is something we have not yet understood" in Higgs physics. 

All this is convincing, and indicates that there is still much we do not know in QFT, regarding the mass of the Higgs, the mechanisms that protects $\lambda$ from scaling, and certainly more.   

But what has all this to do with the question whether in (very) low-energy physics the physical value of the cosmological constant is zero or is small?\footnote%
{One might argue that a mechanism (not yet found) that could protect the cosmological constant from growing too big, would also force it to zero.  But this would simply mean that such particular hypothetical mechanism is not the correct one operating in nature.} 

The question of whether or not there is a cosmological term $\lambda$ in the low-energy classical Einstein equations, is {\em independent} from the question of what is the mechanism that protects this term (zero or small) from being scaled-up to a high scale by radiative corrections.  The first question pertains to low-energy gravitational physics; the second pertains to high-energy particle physics.  The two are independent in the sense that the second question exists independently from the answer to the first. The first has been already answered by observation, as it should: the cosmological term in the Einstein equations does not vanish. The second is open, and has not been changed much by the observations that $\lambda\ne 0$. It is just one of the numerous open problems in high-energy physics.  

We think that the origin of the confusion is that there are two distinct ways of viewing the cosmological term in the action. The first is to assume that this term is {\em nothing else} than the effect of the quantum fluctuations of the vacuum. Namely that $\lambda=0$ in (\ref{scaling}) and the observed acceleration is entirely due to the radiative corrections $\Lambda$ (in the above notation). The second view is that there is a term $\lambda$ in the bare gravitational lagrangian, which might (or might not) be renormalized by radiative corrections. The two points of view are physically different. We think that the common emphasis on the first point of view is wrong.  

In other words, it is a mistake to {\em identify} the cosmological constant $\lambda$ with the zero point energy $\Lambda$ of a QFT, for the same reason one should not {\em a priori} identify the charge of the electron with its radiative corrections. 

If we get confused about this, we make a funny logical mistake. We have an observed physical phenomenon (the accelerated expansion). A simple physical theory explains it (general relativity with nonvanishing $\lambda$).  However, particle physics  claims that it can provide an independent understanding of the phenomenon (a cosmological term entirely produced by vacuum fluctuation). So we discard the simple explanation.  But the new understanding goes wrong quantitatively (by 120 orders of magnitude). Now, every reasonable person would conclude that there is something missing in the particle-physics argument; especially knowing that the argument is already known to be wrong in flat space.  But this is not the conclusion that is usually defended. Rather, it is claimed that what is unacceptable, and needs to be changed is the first simple explanation of the phenomenon!  

There is no known natural way to derive the tiny cosmological constant that plays a role in cosmology from particle physics. And there is no understanding of why this constant is not renormalized to a high value.   But this does not means that there is something mysterious in the cosmological constant itself: it means that there is something we do not understand yet in particle physics. What could this be?

Let us come back to physics and to the simplest reading of the vacuum energy.  The QFT vacuum energy is simple to understand: a harmonic oscillator of frequency $\omega$ has a vacuum energy $E_0=\frac12\hbar \omega$. As a dynamical system, a free field in a box is equivalent to a collection of one oscillator with frequency $\omega_k$ per oscillation mode $k$ of the field.  Hence the quantum field will have a vacuum energy $E_0=\frac12\hbar\sum_k \omega_k$. The modes in a finite box are infinite, and $\omega_k$ grows with $k$, hence $E_0$ diverges.  This is the simplest example of QFT divergence. 

Does this large energy exist for real?  That is, does it have observable effects? In particular: does it act as a source for the gravitational field, as all forms of energy are known to do? Does it have a gravitational mass (and therefore an inertial mass)?  

An effect commonly put forward to support the ``reality" of such a vacuum energy is the Casimir effect.  But the Casimir effect does not reveal the existence of a vacuum energy: it reveals the effect of a ``change" in vacuum energy, and it says nothing about where the zero point value of this energy is. In fact, simple physical arguments indicate that the vacuum energy, by itself, cannot be ``real" in the sense of gravitating: if it did, any empty box containing a quantum field would have a huge mass, and we could not move it with a force, since gravitational mass is also inertial mass.\footnote%
{The relation between the Casimir effect and vacuum energy has been criticized by several authors, on various grounds. For instance by Jaffe, who concludes in \cite{Jaffe} that : ``... no known phenomenon, including the Casimir effect, demonstrates that zero point energies are ``real".}  On physical grounds, vacuum energy does not gravitate. 

A {\em shift} in vacuum energy does gravitate. This is nicely illustrated by an example discussed by Polchinski in \cite{joe}:  the one-loop correction to the electrostatic energy of the nucleus given by the electron loop in Figure 4a. The loop shifts the electrostatic energy by a relative amount of order $\alpha \ln(m_eR_{nuc})/4\pi\sim 10^{-3}$. In different nuclei, electrostatic energy contributes differently to the total mass. For instance it is roughly $10^{-3}$ in aluminum and $3\times10^{-3}$ in platinum. Precision experiments by Dicke (again) and others indicate that aluminum and platinum have the same ratio of gravitational to inertial mass to one part in $10^{12}$ \cite{equiv}. 
Therefore we know to a high precision (one part in $10^6$) that the  {\em shift} in the energy of the nucleus due to the effect shown in Figure 4a does gravitate. On the other hand, we also know empirically that the analogous {\em vacuum} loop in Figure 4b, which is at least of order $O(M_{100\, GeV}^2 m^2_e)$, does {\em not} contribute to $\Lambda$. 
\begin{figure}[h]
\centering
a)\includegraphics[scale=.7]{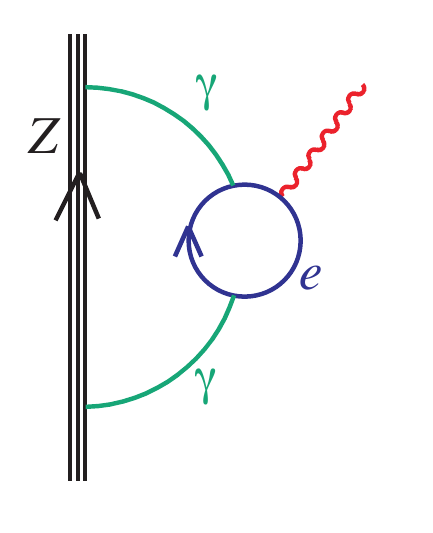}\hskip3em
b)\raisebox{.8cm}{\includegraphics[scale=.7]{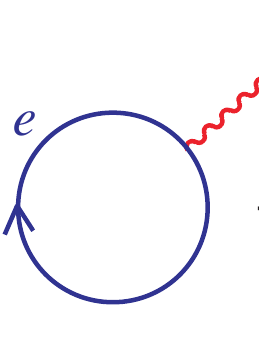}}
\caption{a) The correction to the electrostatic energy of a nucleus given by an electron loop coupled to an external graviton. b) the same  electron loop coupled to an external graviton, in vacuum.}
\end{figure}

Why standard QFT has so much trouble adjusting to this straightforward physical fact? We do not know the answer, but there is a general consideration that may be relevant: in which theoretical context is formulated the argument for large radiative corrections to $\lambda$?  If it is in a context in which we disregard gravity, then a large vacuum energy is physically irrelevant, because the $\lambda$ term in the action (\ref{action}) couples only to the gravitational field $g$, and is invisible if we disregard gravity.  The next option is to formulate it in the context of perturbative QFT including the gravitational field on Minkowski space. Namely to compute (\ref{pi}) in the perturbative expansion (\ref{exp}-\ref{exp2}). But then there is a catch: if  $\lambda$ (or $\lambda_{\rm\scriptscriptstyle  QFT}$) is different from zero, then $(\phi_0,\eta)$ is not a solution of the equations of motion. This is of course because in the presence of a cosmological constant Minkowski space is not anymore a solution of the Einstein field equations: 
\begin{equation}
R_{\mu\nu}[\eta]-\frac12 R\eta_{\mu\nu}[\eta] +\lambda_{\rm\scriptscriptstyle  QFT} \eta_{\mu\nu}= 8\pi G\ T_{\mu\nu}[\eta,\phi_0]
\end{equation}
gives the false result
\begin{equation}
\lambda_{\rm\scriptscriptstyle  QFT} \eta_{\mu\nu} = 0.
\end{equation}
But we cannot rely on a perturbation expansion around a field configuration which is not a solution of the equations of motion. In fact, the tadpole graph in Figure 4b indicates precisely that the gravitational field acquires a vacuum expectation value which is different from the one used in (\ref{exp2}).

Hence we must at the very least move up to the context of QFT on curved spacetimes  \cite{wald}.\footnote{If not to the difficult problem of a perturbative expansion in full background independent quantum gravity \cite{mi}.}  But are we sure that the full perturbation-expansion and renormalization-technique machinery --and especially the intuition-- of flat-space QFT is still valuable on curved spacetimes? It may very well be that the reason there is something unclear in the QFT conclusions about the cosmological constant is not because of some mysterious high-energy effect, but simply because flat space QFT does not apply where a cosmological constant is different from zero, and therefore spacetime is not flat. 

One might object that spacetime is nearly flat locally, and the vacuum energy is a local effect, therefore we can work in the flat-space approximation to compute local quantum corrections to the action. This is fine, but the physical effect of the cosmological constant is not visible in this approximation either: it requires us to go to very large distance, which is precisely where such local approximation fails. 

To trust {\em flat-space} QFT telling us something about the origin or the nature of a term in Einstein equations which implies that spacetime cannot be flat, is a delicate and possibly misleading step.  To argue that a term in Einstein's equations is ``problematic" because flat-space QFT predicts it, but predicts it wrong, seems a {\em non sequitur} to us.  It is saying that a simple explanation is false because an ill-founded alternative explanation gives a wrong answer.

\section{Conclusion}

First, the cosmological constant term is a completely natural part of the Einstein equations. Einstein probably considered it well before thinking about cosmology. His ``blunder" was not to add such a term to the equations: his blunder was to fail to see that the equations, with or without this term, predict expansion.  The term was never seen as unreasonable, or ugly, or a blunder, by the general relativity research community.   It received little attention only because the real value of $\lambda$ is small and its effect was not observed until (as it appears) recently.  

Second, there is no coincidence problem if we consider equiprobability properly, and do not postulate an unreasonably strong cosmological principle, already known to fail.

Third, we do not yet fully understand interacting quantum field theory, its renormalization and its interaction with gravity when spacetime is not Minkowski (that is, in our real universe). But these QFT difficulties have little bearing on the existence of a non vanishing cosmological constant in low-energy physics, because it is a mistake to {\em identify} the cosmological constant with the vacuum energy density. 

As mentioned in the introduction, it is good scientific practice to push the tests of the current theories as far as possible, and to keep studying possible alternatives.  Hence it is necessary to test the $\Lambda$CDM standard model and study alternatives to it, as we do for all physical theories. 

But to claim that dark energy represents a profound mystery, is, in our opinion, nonsense.  ``Dark energy" is just a catch name for the observed acceleration of the universe, which is a phenomenon well described by currently accepted theories, and predicted by these theories, whose intensity is determined by a fundamental constant, now being measured. The measure of the acceleration only determines the value of a constant that was not previously measured.  We have only discovered that a constant that so far (strangely) appeared to be vanishing, in fact is not vanishing. Our universe is full of mystery, but there is no mystery here. 

To claim that ``the greatest mystery of humanity today is the prospect that 75\% of the universe is made up of a substance known as `dark energy' about which we have almost no knowledge at all" is indefensible. 

Why then all the hype about the mystery of the dark energy?  Maybe because great mysteries help getting attention and funding. But a sober and scientifically sound account of what we understand and what we do not understand is preferable for science, on the long run. 

\centerline{---}

We thank S.~Speziale and M.~Smerlak for useful conversations.  After posting this article on the archives, we have received dozens of mails with kind comments and further supporting arguments. We regret there is no place here for thanking everybody individually. The work of E.B. is supported by a Marie Curie Intra-European Fellowship within the 7th European Community Framework Programme.

\end{document}